\documentclass[utf8]{FrontiersinHarvard}
\usepackage{url,hyperref,lineno,microtype,subcaption,soul,color,txfonts,placeins,}
\usepackage{lscape}
\usepackage{lipsum}
\usepackage{subcaption}
\usepackage{hyperref}
\usepackage{graphicx}
\usepackage[percent]{overpic}
\usepackage{mathtools}
\usepackage{enumerate}
\usepackage{ulem}
\usepackage{amsmath}
\usepackage{rotating}
\usepackage{soul}
\usepackage{color}

\linenumbers

\def\keyFont{\fontsize{8}{11}\helveticabold }
\def\firstAuthorLast{Greco E.}

\def\Authors{Emanuele Greco\,$^{1,*}$}
\def\Address{$^{1}$\textrm{INAF-Osservatorio astronomico di Palermo, Palermo, Italy}}
\def\corrAuthor{Emanuele Greco, INAF-Osservatorio Astronomico di Palermo, Piazza del Parlamento 1, 90134, Palermo, Italy}

\def\corrEmail{emanuele.greco@inaf.it}
\nolinenumbers

\begin{document}
\onecolumn
\firstpage{1}

\title[Turbulence in SNRs]{From Shock to Synchrotron: a mini-review on magnetic turbulence in Supernova Remnants}
\author[\firstAuthorLast ]{\Authors} 
\address{} 
\correspondance{} 
\extraAuth{}

\maketitle

\begin{abstract}
Magnetic turbulence plays a crucial role in confining charged particles near the shock front of Supernova Remnants, enabling them to reach energies up to hundreds of TeV through a process known as Diffusive Shock Acceleration (DSA). These high-energy electrons spiral along magnetic field lines, emitting X-ray synchrotron radiation. The launch of the Imaging X-ray Polarimetry Explorer (IXPE) has opened a new window into the study of magnetic fields in SNRs through X-ray polarization measurements. For the first time, IXPE allows us to resolve the polarization degree (PD) and angle (PA) in the X-ray band across different areas of SNRs, offering direct insight into the geometry and coherence of magnetic fields on different scales. In this mini-review, I summarize the key observational results on SNRs obtained with IXPE over the past four years and discuss their implications for our understanding of magnetic turbulence in synchrotron-emitting regions. I also show how we can combine polarization parameters and standard X-ray spectral/imaging analysis to better constrain the structure and scale of magnetic turbulence immediately downstream of the shock and understand the particle acceleration occurring in SNRs.
\keyFont{{ \section{Keywords:} Supernova Remnants, Particle Acceleration, Synchrotron radiation, Magnetic Turbulence, X-ray polarization, X-ray observations}}
\end{abstract}

\section{Introduction}
\label{sec:intro}

Shocks in supernova remnants (SNRs) are widely recognized as prime candidates for the acceleration of Galactic cosmic rays (CRs), primarily through the mechanism of Diffusive Shock Acceleration (DSA, \citealt{fer49,be87}). Within this framework, magnetic turbulence plays a crucial role as it confines charged particles near the shock front, enabling them to repeatedly cross it and gain energy up to PeV energies. The key parameter quantifying this degree of turbulence and the efficiency of particle acceleration is the Bohm factor $\eta = \lambda_{mfp}/r_{g}$, where $\lambda_{mfp}$ is the mean free path of the charged particles and $r_{g}$ their gyroradius. An equivalent expression relates $\eta$ to the ratio between the ordered and turbulent components of the magnetic field $B$ \textbf{(provided that $B > \delta B $, i.e. in the linear regime)}: $\eta=\left< \frac{B}{\delta B} \right>^2$ (see e.g. \citealt{vin20}). For $\eta=1$, in the so-called Bohm diffusion the scattering regime is the most efficient with the particle scattered once per gyro-orbit. The acceleration time scale $t_{acc}$ for relativistic particles under DSA is (\citealt{md01,pmb06,hvb12}):
\begin{equation}
t_{acc} = \eta \frac{E}{3eB} V_s^{-2}
\label{eq:tscale_acc_elec}
\end{equation}
where $E$ is the electron energy and $V_s$ is the shock velocity. TeV electrons move along the magnetic field lines, emit X-ray synchrotron radiation cooling down in a characteristic timescale $t_{\rm loss}$
\begin{equation}
t_{\rm loss} = 12 \left(\frac{B}{1~\textrm{m}G}\right)^{-2} \left(\frac{E}{\mathrm{TeV}}\right)^{-1}~\mathrm{yr}
\label{eq:loss_elec}
\end{equation}

It is often useful to rewrite Eq. \ref{eq:tscale_acc_elec} and Eq. \ref{eq:loss_elec} as functions of the characteristic synchrotron energy $\varepsilon_0$, which can be directly inferred from X-ray spectroscopic analysis:

\begin{equation}
t_{acc} \approx 2~\eta ~\left(\frac{\varepsilon_0}{1~\textrm{keV}}\right)^{0.5} \left(\frac{B}{1~\textrm{mG}}\right)^{-3/2} \left(\frac{V_s}{3000 \, \textrm{km/s}}\right)^{-2} \textrm{yr}; \,\,\,\, t_{\rm loss} \approx 1.5 \left(\frac{\varepsilon_0}{1~\textrm{keV}}\right)^{-0.5}\left(\frac{B}{1~\textrm{m}\textrm{G}}\right)^{-3/2}  ~\textrm{yr}
\label{eq:tscale_acc_cutoff}
\end{equation}

 When the acceleration time scale $t_{acc}$ and the synchrotron loss time scale $t_{\rm loss}$ are comparable, we enter in the so-called loss-limited regime, i.e. the maximum energy of the electrons $E_{max}$ is limited by their losses (\citealt{za07}):

\begin{equation}
E_{\textrm{max}} \approx 10~\eta^{-1/2}~\left(\frac{B}{100~\mu\textrm{G}}\right)^{-1/2}~\left(\frac{V_{\rm s}}{3000~\textrm{km/s}}\right)~~~ \approx ~10 \left(\frac{\varepsilon_0}{1~\textrm{keV}}\right)^{1/2} \left(\frac{B}{100~\mu\textrm{G}}\right)^{-1/2} \textrm{TeV}
\label{eq:maximum_energy_elec}
\end{equation}

where I have used the relation  (\citealt{za07,tuk21,smp24}) for the second part of the equation.

\begin{equation}
\varepsilon_0 = \frac{1.6}{\eta}\left(\frac{V_s}{4000~\textrm{km/s}}\right)^2 \textrm{keV}
\label{eq:epsilon_0_to_eta}
\end{equation}

From Eq. \ref{eq:epsilon_0_to_eta}, one can see that magnetic turbulence, expressed in the form of $\eta$, also affects the spectrum emitted from the charged particles. Equation \ref{eq:epsilon_0_to_eta} offers therefore a direct diagnostic on $\eta$ from X-ray spectroscopic (to measure the cutoff photon energy) and imaging analysis (to measure the shock velocity through proper motion).

Observations with high-resolution X-ray instruments such as Chandra and XMM-Newton (see \citealt{hvb12} for a review) have revealed the presence of narrow synchrotron filaments immediately downstream of the shock. By measuring the thickness of these structures from the images, several studies (e.g., \citealt{vl03}) reported values $ \gtrsim 100 ~\mu G$, much higher than the expected shock-compressed magnetic field of roughly 20 $\mu$G, suggesting strong magnetic field amplification likely induced by CRs-driven instabilities (\citealt{bel04}). The strength of the amplified magnetic field depends on the density of the ambient medium $\rho_0$ and on the shock velocity $V_s$ (\citealt{vin06}) $B^2 \propto \rho_0 V_s^3$, i.e. fast shocks better amplify the magnetic field (\citealt{bla13}) but generate an higher level of magnetic turbulence.

Particle-in-cell (PIC) simulations by \citet{cs14a,cs14b} have shown that self-consistent acceleration of ions can excite magnetic turbulence both upstream and downstream of strong shocks, leading to \(\delta B/B_0 \gtrsim 1\) and a turbulent spectrum characterized by broad-band fluctuations. In these simulations acceleration is most efficient (Bohm-like) for parallel magnetic fields and a radially oriented magnetic structure naturally originates downstream the shock through the Richtmyer-Meshkov instability (RMI, \citealt{ric60,mes69}). Given the small spatial scale in which such simulations are performed, it's hard to immediately translate such structures to the astrophysical scales, but they certainly represent an important hint in understanding the origin of the radially-oriented magnetic field observed in the radio. Magneto-hydrodynamic (MHD) simulations (e.g. \citealt{iso13, bus20}) show that density fluctuations in the ambient medium can seed turbulence via RMI and Rayleigh-Taylor instabilities, which in turn amplify the magnetic field via a turbulent dynamo mechanism. These processes can produce magnetic structures with strengths up to \(\sim 100~\mu\mathrm{G}\) and reproduce the radial-to-tangential magnetic field morphologies observed in young remnants. It is therefore still matter of debate whether the upstream medium plays a role in shaping the downstream magnetic field and whether it could facilitate its amplification and the acceleration of particles.

\subsection{Diagnostic on turbulence through X-ray polarization}
\label{sec:ixpe}
The recent launch, in December 2021, of the Imaging X-ray Polarimetric Explorer (IXPE; \citealt{IXPE}) has opened a new channel to probe turbulence in SNRs: IXPE allows direct, spatially resolved measurements of the X-ray polarization degree (PD) and angle (PA), thus mapping the magnetic-field configuration across different SNRs regions. Since synchrotron radiation is intrinsically linearly polarized - up to $\sim~70\%$ depending on the spectral slope (\citealt{gs65}) - any reduction reflects depolarizing effects such as those relative to magnetic-field amplification. Consequently, PD and PA measurements provide direct insight into the degree of magnetic-field order. Previously, only radio observations were suitable to measure the polarization in SNRs, showing radial fields for young SNRs (e.g. Cas A, \citealt{akr95}; SN 1006, \citealt{rhm13}) and tangential field for more evolved ones (e.g. G156.2+5.7, \citealt{xhs07};  G57.2+0.8, \citealt{ksg18}). However, radio synchrotron radiation is originated by GeV electrons and not by the freshly accelerated TeV electrons responsible for the X-ray emission. Therefore, probing the conditions immediately behind the shock front, where the acceleration is still ongoing, can only be performed by looking at the X-ray polarization data.

Table \ref{tab:ixpe_results} highlights the main features of the sample of SNRs observed, completely or partially, by IXPE so far: Cas A, Tycho, the two limbs of SN1006, the north western region of RX J1713.7-3946 (Vela Jr) and a filament of Vela Jr.

\begin{figure}[!ht]
    \centering
    \includegraphics[width=0.8\linewidth]{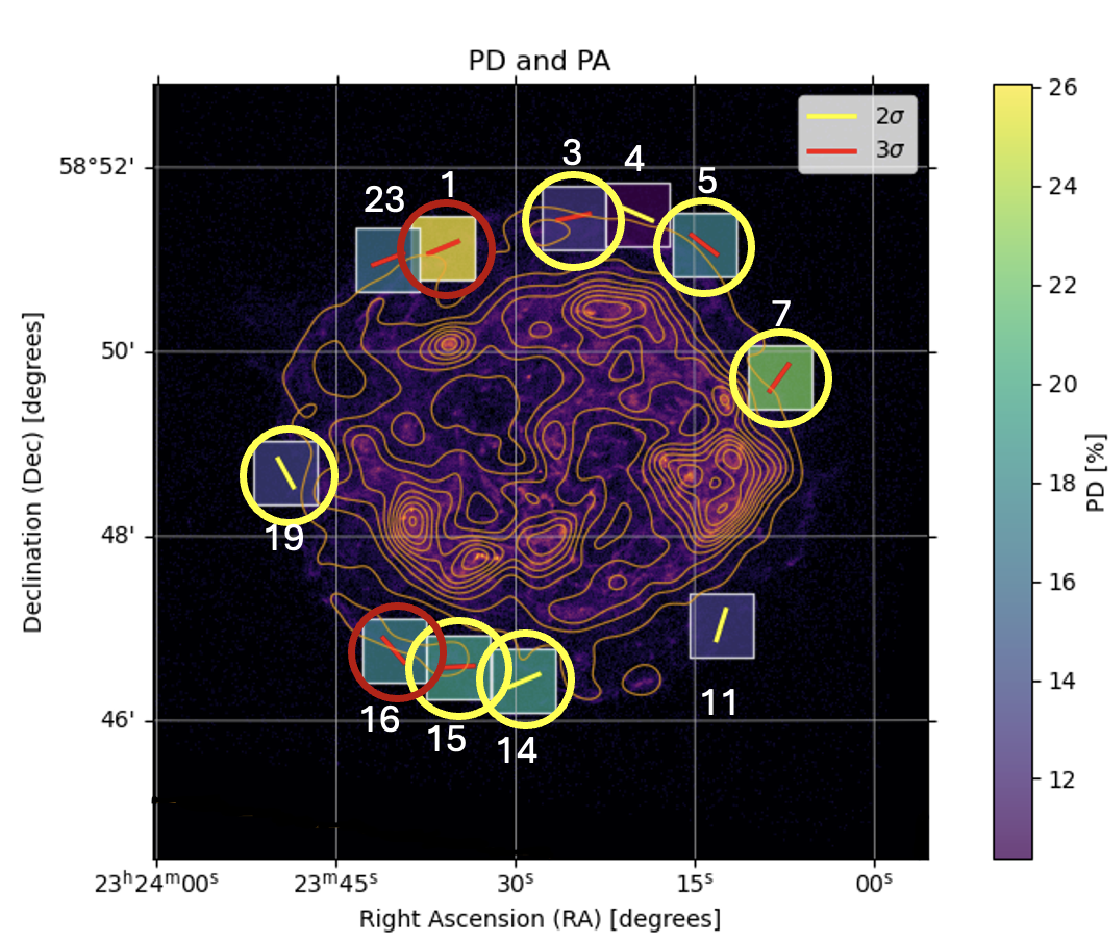}
    \caption{Map of PD reported for Cas A from \citet{mgv25}, overlaid with X-ray polarization vectors. Red and yellow vectors indicate the direction of the X-ray polarization vectors at the 3- and 2-$\sigma$ confidence level, respectively. Red and yellow circles mark regions detected at the 3- and 2-$\sigma$ confidence level with the approach used by \citet{vpf22}, respectively. Superimposed in orange are the 3-6 keV Chandra contours.}
    \label{fig:pol_casa}
\end{figure}

\subsubsection{Cas A}
Cas A, a $\sim 350$ yr old core-collapse SNR, was the first science target observed by IXPE. Its forward shock velocity is of the order of 5000 km/s and shows a very low overall polarization level in the radio band of $\sim 5\%$ (e.g. \citealt{akr95}) and a radial magnetic field (\citealt{ros70}). Analysis performed by \citet{vpf22} retrieved a value of 2-5\% in the PD for the whole shell and forward shock regions, lower than that observed in the radio, with a magnetic field radially oriented. \citet{mgv25} re-analyzed the dataset adopting a different approach based on spectropolarimetric analysis: they fitted the IXPE spectra extracted from boxes across the whole shell by properly accounting for the thermal emission through the analysis of Chandra observations. They found PD values as high as 26\%, at more than the 3$\sigma$ confidence level, highlighting that within relatively big and inhomogeneous regions it is possible to have an ordered magnetic field locally (Fig. \ref{fig:pol_casa}).

\subsubsection{Tycho}
\begin{figure}[!hb]
    \centering
    \includegraphics[width=0.8\linewidth]{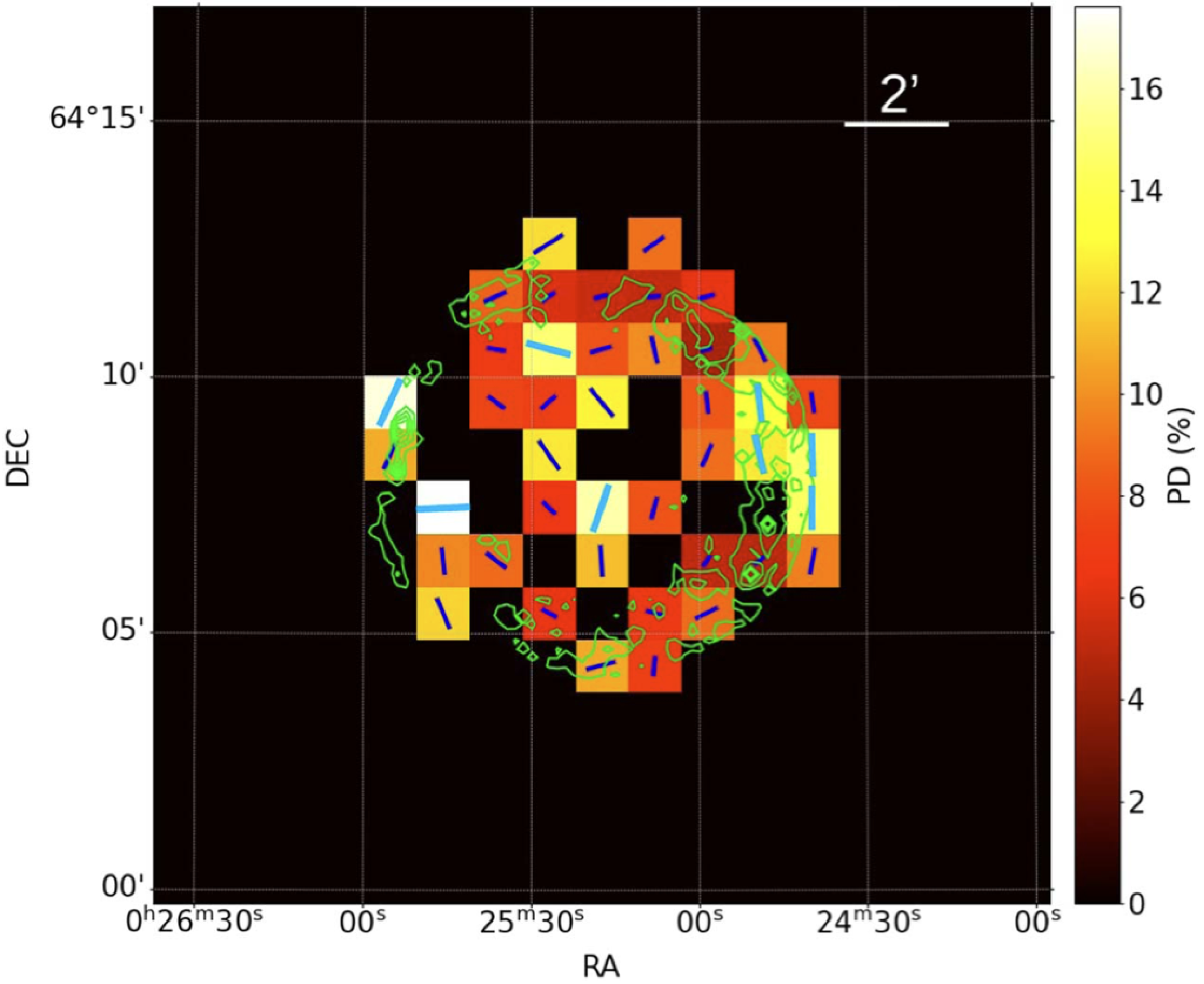}
    \caption{Map of PD reported for Tycho's SNR from \citet{fsp23}. Blue vectors represent the direction of the polarization and their length is proportional to the degree of polarization. The thicker cyan bars mark the pixels with significance higher than 2$\sigma$. Superimposed in green are the 4–6 keV Chandra contours.}
    \label{fig:pol_tycho}
\end{figure}

The historical Tycho's SNR, exploded in 1572 is a Type Ia SNR, characterized by shock velocities of around 4000 km/s (\citealt{wch16}). Significant radio polarization was found only in the outer rim at the $\sim 7\%$ level and the magnetic field was found to be radially oriented (\citealt{dvs91}). Tycho is characterised by thin X-ray synchrotron structures in the west, known as "stripes" and first identified by \citet{ehb11} and characterised by variations on scale of few years (\citealt{mtu20}). The overall X-ray polarization was found to be $9\%$ in the whole SNR and $12\%$ if considering only the shell, where the synchrotron radiation is dominating (\citealt{fsp23}, see Fig. \ref{fig:pol_tycho}). The stripes did not show a remarkably higher or lower PD with respect to the other regions.

\begin{figure}[!ht]
    \centering
    \includegraphics[width=0.43\linewidth]{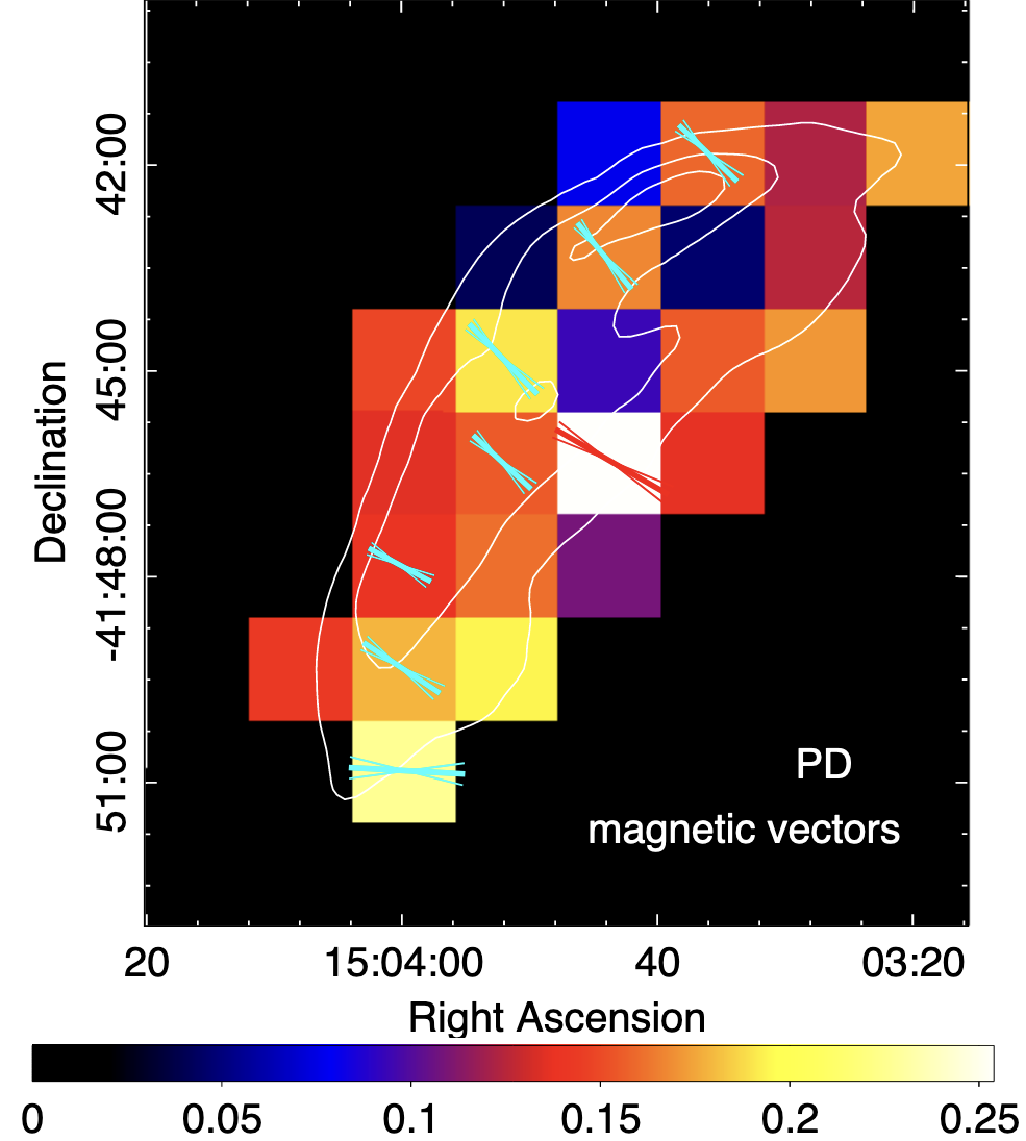}
    \includegraphics[width=0.46\linewidth]{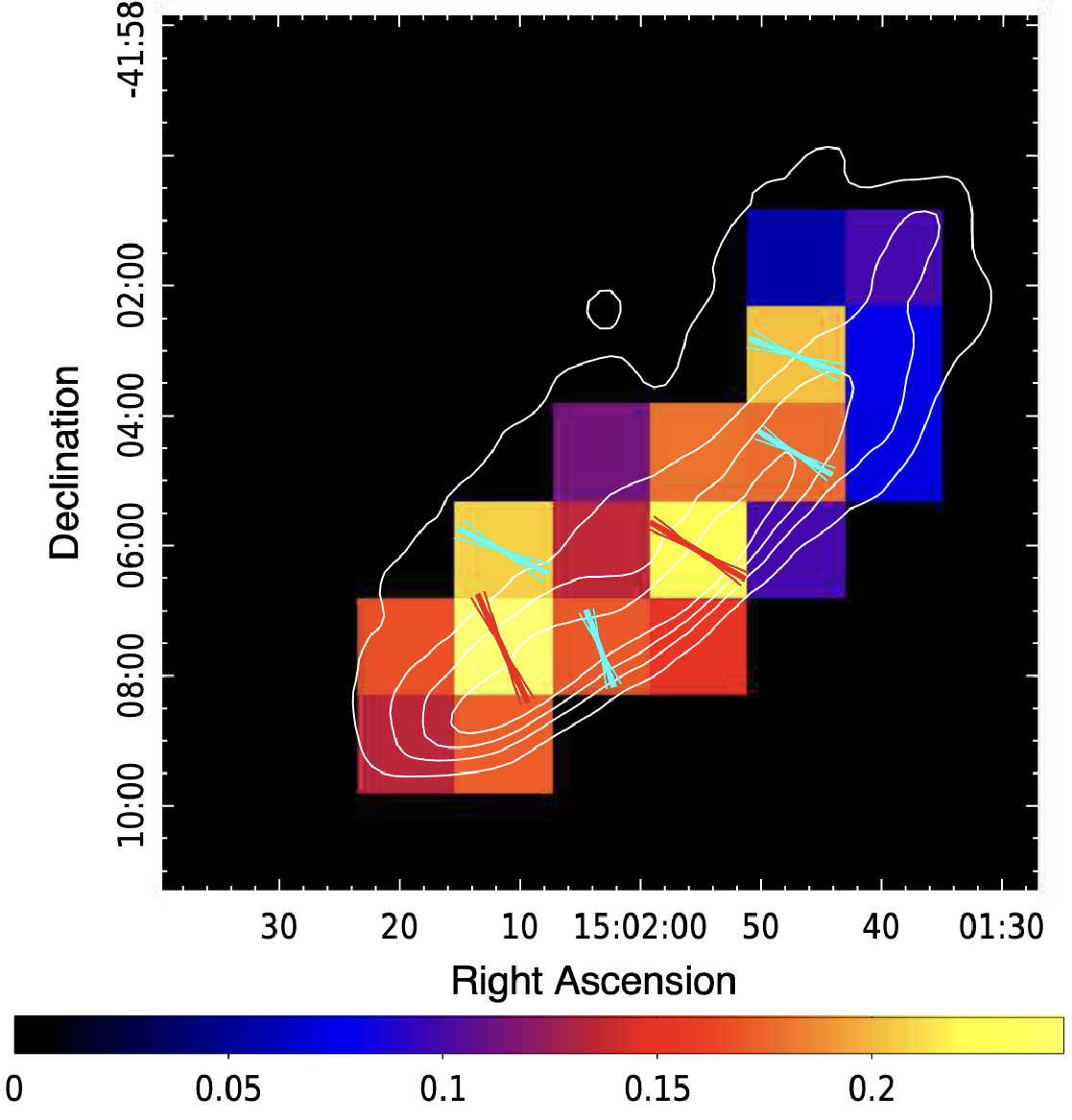}
    \caption{Map of PD reported for the NE and SW limbs of SN 1006 from \citet{zpf23} and \citet{zsp25}, in the left and right panels respectively.  Vectors represent the direction of the magnetic field and their $1\sigma$ uncertainties, with blue corresponding to significance of 2--3$\sigma$ and $> 3\sigma$, respectively. White contours show the Stokes I levels.}
    \label{fig:pol_sn1006}
\end{figure}
\subsubsection{SN 1006}
The remnant of the historical SN 1006 is popular for being the first one for which X-ray synchrotron radiation was observed in its limbs (\citealt{kpg95}). IXPE observed both the NorthEastern (NE, \citealt{zpf23}) and SouthWestern (SW, \citealt{zsp25}) limbs finding for both the limbs a radially-oriented magnetic field and an X-ray PD of $\sim$ 22\%, the highest reported for SNRs. However, the distribution of the PD is quite different between the two limbs (Fig. \ref{fig:pol_sn1006}). In the NE the PD is quite homogenous across the whole area, whereas in the SW the minimum value of PD is found in a region known to interact with an HI cloud (\citealt{mad14}) and peaks up to $\sim 40\%$ where there is no interaction. This is suggestive of environment-dependent magnetic turbulence and magnetic field amplification.

\subsubsection{RX J1713.7-3946}
RX J1713.7-3946 (hereafter RX J1713) is a close ($\sim 1$ kpc) and large SNR having a diameter of roughly 1 degree. Its X-ray emission is completely synchrotron-dominated, with the only exception for a detection of shocked ejecta in the inner area (\citealt{kat15}). 
At odds with the other SNRs discussed so far, IXPE observations of the NW area of RX J1713 indicated a tangential magnetic field and an overall polarization degree of roughly $12\%$ (\citealt{fpb24}, left panel in Fig. \ref{fig:pol_rxj-vela}). This could be either due to the higher age of the SNR, in analogy with what observed for evolved SNRs in the radio band, or to the different environment in which the shock is expanding, characterized by dense molecular clouds.

\subsubsection{Vela Jr}
\begin{figure}[!ht]
    \centering
    \includegraphics[width=0.4\linewidth]{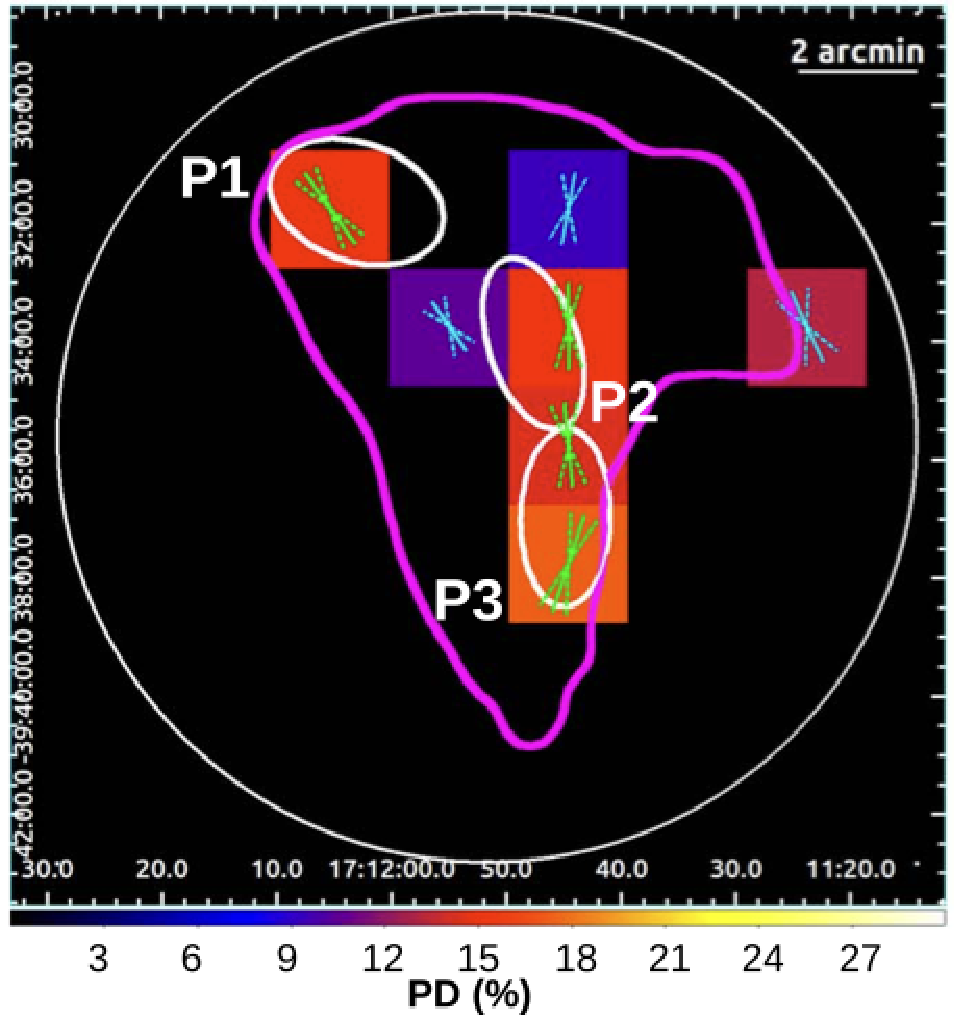}
    \includegraphics[width=0.5\linewidth]{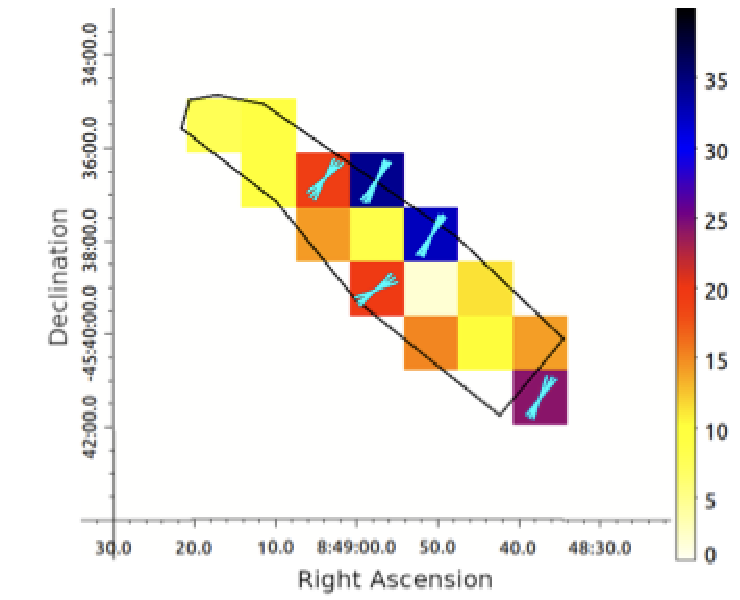}
    \caption{\textit{Left panel. }PD and PA values reported for the NW area of RX J1713 from \citet{fpb24}. Cyan and green vectors mark the magnetic field direction at the 2- and 3-$\sigma$ significance level, respectively. Dashed vectors indicate the $2\sigma$ uncertainty on the direction. The magenta line indicates the 2-5 keV IXPE contours. \textit{Right panel.} Same as left panel but for the NW rim of Vela Jr from \citet{pyf24}. The black line encloses all the pixels with significant X-ray flux.}
    \label{fig:pol_rxj-vela}
\end{figure}

Vela Jr shares several features with RX J1713, being dominated by X-ray nonthermal emission, having an angular size of roughly 2 degrees and being roughly $2000$ yr old (\citealt{ktm08}). IXPE targeted the NW rim of Vela Jr (\citealt{pyf24}), which is the brightest in the X-rays, providing similar results to those obtained for RX J1713, with a tangential magnetic field and a PD $\sim 16\%$ (right panel in Fig. \ref{fig:pol_rxj-vela}).

\subsection{Diagnostic on turbulence through X-ray spectra}
\label{sec:spectra}

\begin{figure}
    \centering
    \includegraphics[width=0.8\linewidth]{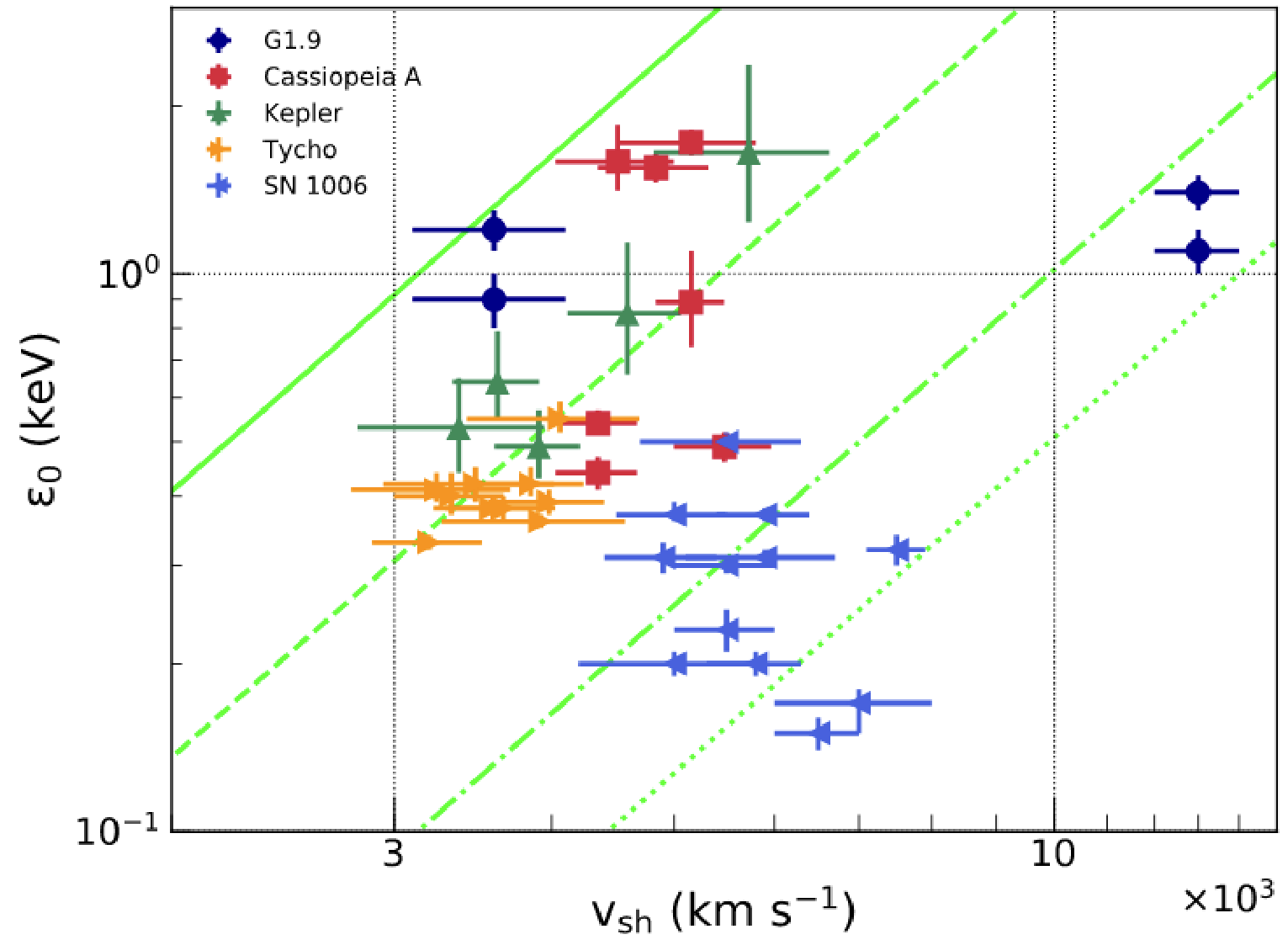}
    \caption{Shock velocity $v_{sh}$ vs cutoff parameter $\varepsilon_0$ plot from \citet{tuk21} for different regions among the SNRs G1.9+0.3, Cas A, Kepler, Tycho and SN 1006. Green solid, dashed, dashed-dotted and dotted lines indicate $\eta$ of 1, 3, 10 and 20, respectively.}
    \label{fig:eta_tsuji}
\end{figure}

X-ray spectral analysis by itself is a powerful diagnostic tool to estimate the degree of turbulence in the plasma emitting X-ray synchrotron radiation, mainly through the estimate of the Bohm factor $\eta$. One of the most straightforward examples is provided by \citet{tuk21}, who systematically applied Eq. \ref{eq:epsilon_0_to_eta} to a sample of SNRs characterized by nonthermal emission, to measure the value of $\eta$ (see Fig. \ref{fig:eta_tsuji}.
They found that the Tycho's and Kepler's SNR nicely match the theoretical prediction, indicating a constant $\eta$ across the various region considered. On the other hand, Cas A showed very different $\eta$ values across regions with similar shock velocities. SN 1006 presented lower $\eta$ values in correspondence of the nonthermal limbs, as expected given the configuration of the magnetic field and the efficiency of acceleration in the quasi-parallel scenario (\citealt{cs14a}). In their work, \citet{tuk21} could not perform spatially resolved analysis for other SNRs and reported average values of $\eta \sim 1$ both for RX J1713 and Vela JR.
Regarding Kepler's SNR, \citet{smb22} reported two different acceleration regimes between the north, site of interaction with a clump of CSM (\citealt{rbh07}), and the south, where the shock wave is expanding freely. In the north they report values $\eta \sim 1$, while in the south $\eta \sim 6$.

Despite its crucial role in the DSA paradigm and the low PD observed, prevailing models used to fit X-ray nonthermal spectra of SNRs (e.g., \texttt{SRcut} by \citet{rk99}, the loss-limited model by \citet{za07}, or a simple phenomenological \texttt{power-law}) are calculated assuming an homogeneous magnetic field. The resulting analytical expressions are typically power-law with a cutoff, with the shape of the cutoff depending mainly on whether the maximum energy of the electrons is time-limited or loss-limited. Therefore, it is natural to ask what the emitted spectrum would look like by considering a turbulent magnetic field. This has been explored theoretically by \citet{tf87} and \citet{kak13}, who labeled "jitter radiation" this emission process in which the electrons are sensitive also to the magnetic field's turbulent component. The resulting jitter radiation photon spectrum emitted from a power-law distribution of electrons can be described as a broken power-law having a smooth break between the regime dominated by standard synchrotron radiation (at lower energy) and that dominated by the jitter component (at higher energy). Interestingly, despite using different approximation and assumptions, both \citet{tf87} and \citet{kak13} agree on two striking features of jitter radiation: i) the slope of the spectrum at high energies is directly linked to the turbulence spectrum; ii) the maximum energy to which the spectrum extends is proportional to minimum scale of the turbulence. Jitter radiation is therefore a powerful tool that could provide direct diagnostics on turbulence parameters otherwise inaccessible for astrophysical sources. \citet{gve23} applied the jitter paradigm to the SNR Cas A, finding that it describes the X-ray broadband nonthermal spectra better than any standard cutoff model. They inferred an index in the turbulence spectrum $\nu_B$ of 2-2.4, depending on the region considered, and an upper limit on the turbulence minimum scale of $\sim 70$ km. \textbf{It is worth noting that in order for jitter radiation to be at work, the minimum scale of turbulence has also to be lower than the synchrotron formation length, since at larger scales the standard equations for synchrotron emission from electrons embedded in an homogeneous magnetic field would still be valid locally. Remarkably, for Cas A such criterion is satisfied, as the typical value for the synchrotron formation length is of $\sim 10^4$ km.}

\section{Discussion}
In section \ref{sec:intro} of this mini-review I briefly recalled the main recent results strictly related to the degree of turbulence and acceleration efficiency in the X-ray nonthermal emission of SNRs. The first clear takeaway is that acceleration efficiency - and, therefore, magnetic turbulence - is very high in most X-ray synchrotron-emitting SNR, with the case of Vela Jr and RX J1713 showcasing the lowest values of Bohm factor $\eta$. I also showed that all the considered SNRs show PD much lower than the intrinsic limit, in line with the requirement of high magnetic turbulence for an highly efficient acceleration process. Some of these SNR present different acceleration regimes across different areas, depending on the conditions of the ambient environment and anisotropies. This is particularly true for Cas A and Kepler's SNR. The former is the SNR with the lowest X-ray polarization highlighting that it's evolving in a dense and inhomogeneous medium, relic of the progenitor star. Similarly, in the north the shock wave of Kepler's is interacting with a shell of CSM, and despite not being observed by IXPE yet, it shows an higher radio polarization in the south with respect to the north (\citealt{dkr02}). On the other hand, SNRs such as SN 1006, RX J1713 and Vela Jr, which evolve in rarefied environment, show polarization levels higher than those observed in Cas A. These results indicate that while most of the turbulence responsible for the depolarization of the synchrotron radiation is intrinsically developed by the DSA mechanism, a secondary but definitely important role is also played by the preexisting medium. We are currently short of detailed simulations which investigate the acceleration process occurring in a shock expanding through a turbulent environment, and it is likely that combined effects of the instabilities intrinsically generated and the extrinsic inhomogeneities in the ambient media could help us improve our understanding of the acceleration process and our interpretation of the PD values across different areas in SNRs.

It is important to notice that there is not necessarily a 1-to-1 relationship between the bohm factor $\eta$ and the PD measured by IXPE. As also pointed out by \citet{pyf24}, $\eta$ quantifies the diffusion coefficient along the shock normal (\citealt{za07}), while PD and PA depends on the distribution of the projected magnetic field onto the plane of the sky. This is a critical point for SNRs such as RX J1713 and Vela Jr, showing a tangential magnetic field, a relatively high polarization of $10\%$ but a very efficient acceleration regime, having $\eta \approx 1$. The observed tangential polarization pattern indicates that only this component of the magnetic field retains a coherent structure, while the radial component is highly turbulent. Such radial disorder is exactly what one expects in a Bohm-like regime ($\eta \sim$ 1), as reported for RX J1713 and Vela Jr and indicates a strongly anisotropic turbulence.

\textbf{Turbulence does not always imply depolarization, especially in case in which there is a strong anisotropy, as has been shown for Tycho's SNR by \citet{bou24}.} Anisotropic turbulence can also easily lead to ordered magnetic field structures that can increase the polarization level locally. For example, anisotropic turbulence may locally enhance/decrease the acceleration efficiency, inducing lower/higher values of $\eta$ and of the radial polarization. This might be the case for some of the region in Cas A analysed by \citet{vpf22} and \citet{mgv25}, who identified regions with PD levels even a factor $\approx$ 6 higher than the average on the whole shell. Naturally, it is not obvious to disentangle this kind of effects from the depolarization due to the mixing of the emission from different regions. The creation of coherent structures due to magnetic turbulence might sound counterintuitive, but it could explain, for example, also the flickering of X-ray hot spots reported by \citet{uat07} for RX J1713: fluctuations might generate extremely high ($\sim$ mG) magnetic field in localized regions. One would expect that radiation stemming from these spots is highly unpolarized, but this may not be entirely true because of, again, the putative different orientation on the magnetic field with respect to the shock normal. IXPE has not the capability to resolve arcsecond-like structures, but an hint on this could be achieved by cross relating the regions characterized by lower magnetic field (i.e. less amplified and therefore less turbulent) with those having higher PD values. Remembering that magnetic field is better amplified in faster shocks, the velocity of the shock could be used as first key parameter to identify the most promising regions, excluding cases with clear interaction with significantly denser media.

In Sect. \ref{sec:intro} I also described a new potential spectroscopic diagnostic tool on turbulence, based on the jitter radiation paradigm. The only application on SNR was performed by \citet{gve23} on Cas A. They found a turbulence spectrum steeper than the typical Kolmogorov value of $\nu_B = 5/3$, suggesting that different form of instabilities might be at work. Surprisingly, the turbulence scales inferred are much lower than those considered for magnetic turbulence responsible for acceleration, of the order of $\sim 100 $ km. This point is not necessarily an issue since in the jitter framework the inclusion of a turbulent component in the magnetic field only affects the emission of photons and not the acceleration process. It might be more questionable, and future studies are needed in this sense, whether turbulence can actually survive at this scale. 
w
In general, the jitter radiation process needs to be tested and validated in a more systematic way for several other SNRs. In this framework, a decisive signature would be a systematic correlation between the PD measured by IXPE and the jitter break energy: a higher break energy would indicate that the intrinsically polarized synchrotron component dominates in the 4–6 keV band, whereas a lower break energy would imply that the intrinsically unpolarized jitter component already prevails in this range.

\section{Conclusions}

In this mini-review I summarized some of the most relevant recent results inferred from X-ray polarimetric and spectral analysis relative to magnetic turbulence in SNRs. The current picture clearly shows that turbulence is a key ingredient in the acceleration and X-ray emission mechanism in SNRs. Thanks to the IXPE telescope it is now possible to measure the PD and PA in several SNRs. These unprecedented information coupled with the results obtained on the curvature of the X-ray nonthermal spectra show that every synchrotron-emitting SNR is characterized by turbulence, mainly self-generated by the expanding shock. However, also the configuration of the pre-existing medium seems to play a significant role. Additional observations and numerical simulations are needed to better understand to what extent the preexisting medium affects the acceleration and emission processes.

\begin{table*}[ht]
\centering
\caption{Sample of SNRs observed by IXPE.}
\begin{tabular}{lccccccl}
\hline\hline
\textbf{SNR} & \textbf{Age (yr)} & \boldmath$B~(\mu\text{G})$ & \textbf{PD} & \boldmath$V_s~$\unboldmath \textbf{(km/s)} & \boldmath$B$ \textbf{orientation} &  \textbf{References} \\
\hline
Cas A (whole)& $\sim 350$& $\sim 250$ & $\sim 2.5\%$ & $\sim 5000$& radial      & \citet{vpf22} \\
Cas A (regions)& $\sim 350$& $\sim 250$ & $10-26$\% & $\sim 5000$& radial      & \citet{mgv25}\\
Tycho’s SNR& 453& $\sim 200$  & $\sim 9\%$& $\sim 4000$& radial& \citet{fsp23}\\
SN 1006 (NE)& 1019& $\sim 80$& $\sim 22\%$& $\sim 5000$ &radial&  \citet{zpf23}\\
SN 1006 (SW)& 1019& $\sim 80$& $20-40\%$ & $\sim 5000$ &radial&  \citet{zsp25}\\
RX J1713 (NW)& $\sim 1500$& $\sim 20$& 26--30\%& $\sim 4000$ & tangential &  \citet{fpb24}\\
Vela Jr (NW) & $\sim 2000$ & $\sim 10$ & 10--20\%&  $\gtrsim 3000$& tangential  & \citet{pyf24}\\
\hline
\label{tab:ixpe_results}
\end{tabular}
\end{table*}


\section*{Acknowledgments}
EG acknowledge discussion and collaboration with colleagues M. Miceli, S. Orlando, J. Vink, A. Mercuri, S. Perri, D. Caprioli. EG acknowledges support from the INAF Minigrant RSN4 “Investigating magnetic turbulence in young Supernova Remnants through X-ray observations”.



\bibliographystyle{Frontiers-Harvard}
\bibliography{References}

\begin{thebibliography}{50}
\providecommand{\natexlab}[1]{#1}
\expandafter\ifx\csname urlstyle\endcsname\relax
  \providecommand{\doi}[1]{doi:\discretionary{}{}{}#1}\else
  \providecommand{\doi}{doi:\discretionary{}{}{}\begingroup \urlstyle{rm}\Url}\fi
\providecommand{\selectlanguage}[1]{\relax}
\providecommand{\bibAnnoteFile}[1]{%
  \IfFileExists{#1}{\begin{quotation}\noindent\textsc{Key:} #1\\
  \textsc{Annotation:}\ \input{#1}\end{quotation}}{}}
\providecommand{\bibAnnote}[2]{%
  \begin{quotation}\noindent\textsc{Key:} #1\\
  \textsc{Annotation:}\ #2\end{quotation}}

\bibitem[{{Anderson} et~al.(1995){Anderson}, {Keohane}, and {Rudnick}}]{akr95}
{Anderson}, M.~C., {Keohane}, J.~W., and {Rudnick}, L. (1995).
\newblock {The Polarization and Depolarization of Radio Emission from Supernova Remnant Cassiopeia A}.
\newblock \emph{\apj} 441, 300.
\newblock \doi{10.1086/175356}
\bibAnnoteFile{akr95}

\bibitem[{{Bell}(2004)}]{bel04}
{Bell}, A.~R. (2004).
\newblock {Turbulent amplification of magnetic field and diffusive shock acceleration of cosmic rays}.
\newblock \emph{\mnras} 353, 550--558.
\newblock \doi{10.1111/j.1365-2966.2004.08097.x}
\bibAnnoteFile{bel04}

\bibitem[{{Blandford} and {Eichler}(1987)}]{be87}
{Blandford}, R. and {Eichler}, D. (1987).
\newblock {Particle Acceleration at Astrophysical Shocks - a Theory of Cosmic-Ray Origin}.
\newblock \emph{\physrep} 154, 1--+
\bibAnnoteFile{be87}

\bibitem[{{Blasi}(2013)}]{bla13}
{Blasi}, P. (2013).
\newblock {The origin of galactic cosmic rays}.
\newblock \emph{\aapr} 21, 70.
\newblock \doi{10.1007/s00159-013-0070-7}
\bibAnnoteFile{bla13}

\bibitem[{{Bykov} et~al.(2024){Bykov}, {Osipov}, {Uvarov}, {Ellison}, and {Slane}}]{bou24}
{Bykov}, A.~M., {Osipov}, S.~M., {Uvarov}, Y.~A., {Ellison}, D.~C., and {Slane}, P. (2024).
\newblock {X-ray polarization: A view deep inside cosmic ray driven turbulence and particle acceleration in supernova remnants}.
\newblock \emph{\prd} 110, 023041.
\newblock \doi{10.1103/PhysRevD.110.023041}
\bibAnnoteFile{bou24}

\bibitem[{{Bykov} et~al.(2020){Bykov}, {Uvarov}, {Slane}, and {Ellison}}]{bus20}
{Bykov}, A.~M., {Uvarov}, Y.~A., {Slane}, P., and {Ellison}, D.~C. (2020).
\newblock {Uncovering Magnetic Turbulence in Young Supernova Remnants with Polarized X-Ray Imaging}.
\newblock \emph{\apj} 899, 142.
\newblock \doi{10.3847/1538-4357/aba960}
\bibAnnoteFile{bus20}

\bibitem[{{Caprioli} and {Spitkovsky}(2014{\natexlab{a}})}]{cs14a}
{Caprioli}, D. and {Spitkovsky}, A. (2014{\natexlab{a}}).
\newblock {Simulations of Ion Acceleration at Non-relativistic Shocks. I. Acceleration Efficiency}.
\newblock \emph{\apj} 783, 91.
\newblock \doi{10.1088/0004-637X/783/2/91}
\bibAnnoteFile{cs14a}

\bibitem[{{Caprioli} and {Spitkovsky}(2014{\natexlab{b}})}]{cs14b}
{Caprioli}, D. and {Spitkovsky}, A. (2014{\natexlab{b}}).
\newblock {Simulations of Ion Acceleration at Non-relativistic Shocks. II. Magnetic Field Amplification}.
\newblock \emph{\apj} 794, 46.
\newblock \doi{10.1088/0004-637X/794/1/46}
\bibAnnoteFile{cs14b}

\bibitem[{{DeLaney} et~al.(2002){DeLaney}, {Koralesky}, {Rudnick}, and {Dickel}}]{dkr02}
{DeLaney}, T., {Koralesky}, B., {Rudnick}, L., and {Dickel}, J.~R. (2002).
\newblock {Radio Spectral Index Variations and Physical Conditions in Kepler's Supernova Remnant}.
\newblock \emph{\apj} 580, 914--927.
\newblock \doi{10.1086/343787}
\bibAnnoteFile{dkr02}

\bibitem[{{Dickel} et~al.(1991){Dickel}, {van Breugel}, and {Strom}}]{dvs91}
{Dickel}, J.~R., {van Breugel}, W.~J.~M., and {Strom}, R.~G. (1991).
\newblock {Radio Structure of the Remnant of Tycho's Supernova (SN 1572)}.
\newblock \emph{\aj} 101, 2151.
\newblock \doi{10.1086/115837}
\bibAnnoteFile{dvs91}

\bibitem[{{Eriksen} et~al.(2011){Eriksen}, {Hughes}, {Badenes}, {Fesen}, {Ghavamian}, {Moffett} et~al.}]{ehb11}
{Eriksen}, K.~A., {Hughes}, J.~P., {Badenes}, C., {Fesen}, R., {Ghavamian}, P., {Moffett}, D., et~al. (2011).
\newblock {Evidence for Particle Acceleration to the Knee of the Cosmic Ray Spectrum in Tycho's Supernova Remnant}.
\newblock \emph{\apjl} 728, L28.
\newblock \doi{10.1088/2041-8205/728/2/L28}
\bibAnnoteFile{ehb11}

\bibitem[{{Fermi}(1949)}]{fer49}
{Fermi}, E. (1949).
\newblock {On the Origin of the Cosmic Radiation}.
\newblock \emph{Physical Review} 75, 1169--1174.
\newblock \doi{10.1103/PhysRev.75.1169}
\bibAnnoteFile{fer49}

\bibitem[{{Ferrazzoli} et~al.(2024){Ferrazzoli}, {Prokhorov}, {Bucciantini}, {Slane}, {Vink}, {Cardillo} et~al.}]{fpb24}
{Ferrazzoli}, R., {Prokhorov}, D., {Bucciantini}, N., {Slane}, P., {Vink}, J., {Cardillo}, M., et~al. (2024).
\newblock {Discovery of a Shock-compressed Magnetic Field in the Northwestern Rim of the Young Supernova Remnant RX J1713.7{\textendash}3946 with X-Ray Polarimetry}.
\newblock \emph{\apjl} 967, L38.
\newblock \doi{10.3847/2041-8213/ad4a68}
\bibAnnoteFile{fpb24}

\bibitem[{{Ferrazzoli} et~al.(2023){Ferrazzoli}, {Slane}, {Prokhorov}, {Zhou}, {Vink}, {Bucciantini} et~al.}]{fsp23}
{Ferrazzoli}, R., {Slane}, P., {Prokhorov}, D., {Zhou}, P., {Vink}, J., {Bucciantini}, N., et~al. (2023).
\newblock {X-Ray Polarimetry Reveals the Magnetic-field Topology on Sub-parsec Scales in Tycho's Supernova Remnant}.
\newblock \emph{\apj} 945, 52.
\newblock \doi{10.3847/1538-4357/acb496}
\bibAnnoteFile{fsp23}

\bibitem[{{Ginzburg} and {Syrovatskii}(1965)}]{gs65}
{Ginzburg}, V.~L. and {Syrovatskii}, S.~I. (1965).
\newblock {Cosmic Magnetobremsstrahlung (synchrotron Radiation)}.
\newblock \emph{\araa} 3, 297.
\newblock \doi{10.1146/annurev.aa.03.090165.001501}
\bibAnnoteFile{gs65}

\bibitem[{{Greco} et~al.(2023){Greco}, {Vink}, {Ellien}, and {Ferrigno}}]{gve23}
{Greco}, E., {Vink}, J., {Ellien}, A., and {Ferrigno}, C. (2023).
\newblock {Jitter Radiation as an Alternative Mechanism for the Nonthermal X-Ray Emission of Cassiopeia A}.
\newblock \emph{\apj} 956, 116.
\newblock \doi{10.3847/1538-4357/acf567}
\bibAnnoteFile{gve23}

\bibitem[{{Helder} et~al.(2012){Helder}, {Vink}, {Bykov}, {Ohira}, {Raymond}, and {Terrier}}]{hvb12}
{Helder}, E.~A., {Vink}, J., {Bykov}, A.~M., {Ohira}, Y., {Raymond}, J.~C., and {Terrier}, R. (2012).
\newblock {Observational Signatures of Particle Acceleration in Supernova Remnants}.
\newblock \emph{\ssr} 173, 369--431.
\newblock \doi{10.1007/s11214-012-9919-8}
\bibAnnoteFile{hvb12}

\bibitem[{{Inoue} et~al.(2013){Inoue}, {Shimoda}, {Ohira}, and {Yamazaki}}]{iso13}
{Inoue}, T., {Shimoda}, J., {Ohira}, Y., and {Yamazaki}, R. (2013).
\newblock {The Origin of Radially Aligned Magnetic Fields in Young Supernova Remnants}.
\newblock \emph{\apjl} 772, L20.
\newblock \doi{10.1088/2041-8205/772/2/L20}
\bibAnnoteFile{iso13}

\bibitem[{{Katsuda} et~al.(2015){Katsuda}, {Acero}, {Tominaga}, {Fukui}, {Hiraga}, {Koyama} et~al.}]{kat15}
{Katsuda}, S., {Acero}, F., {Tominaga}, N., {Fukui}, Y., {Hiraga}, J.~S., {Koyama}, K., et~al. (2015).
\newblock {Evidence for Thermal X-Ray Line Emission from the Synchrotron-dominated Supernova Remnant RX J1713.7-3946}.
\newblock \emph{\apj} 814, 29.
\newblock \doi{10.1088/0004-637X/814/1/29}
\bibAnnoteFile{kat15}

\bibitem[{{Katsuda} et~al.(2008){Katsuda}, {Tsunemi}, and {Mori}}]{ktm08}
{Katsuda}, S., {Tsunemi}, H., and {Mori}, K. (2008).
\newblock {The Slow X-Ray Expansion of the Northwestern Rim of the Supernova Remnant RX J0852.0-4622}.
\newblock \emph{\apjl} 678, L35.
\newblock \doi{10.1086/588499}
\bibAnnoteFile{ktm08}

\bibitem[{{Kelner} et~al.(2013){Kelner}, {Aharonian}, and {Khangulyan}}]{kak13}
{Kelner}, S.~R., {Aharonian}, F.~A., and {Khangulyan}, D. (2013).
\newblock {On the Jitter Radiation}.
\newblock \emph{\apj} 774, 61.
\newblock \doi{10.1088/0004-637X/774/1/61}
\bibAnnoteFile{kak13}

\bibitem[{{Kothes} et~al.(2018){Kothes}, {Sun}, {Gaensler}, and {Reich}}]{ksg18}
{Kothes}, R., {Sun}, X., {Gaensler}, B., and {Reich}, W. (2018).
\newblock {A Radio Continuum and Polarization Study of SNR G57.2+0.8 Associated with Magnetar SGR 1935+2154}.
\newblock \emph{\apj} 852, 54.
\newblock \doi{10.3847/1538-4357/aa9e89}
\bibAnnoteFile{ksg18}

\bibitem[{{Koyama} et~al.(1995){Koyama}, {Petre}, {Gotthelf}, {Hwang}, {Matsura}, {Ozaki} et~al.}]{kpg95}
{Koyama}, K., {Petre}, R., {Gotthelf}, E.~V., {Hwang}, U., {Matsura}, M., {Ozaki}, M., et~al. (1995).
\newblock Evidence for shock acceleration of high-energy electrons in the supernova remnant sn:1006.
\newblock \emph{\nat} 378, 255
\bibAnnoteFile{kpg95}

\bibitem[{{Malkov} and {O'C Drury}(2001)}]{md01}
{Malkov}, M.~A. and {O'C Drury}, L. (2001).
\newblock {Nonlinear theory of diffusive acceleration of particles by shock waves}.
\newblock \emph{Reports on Progress in Physics} 64, 429--481.
\newblock \doi{10.1088/0034-4885/64/4/201}
\bibAnnoteFile{md01}

\bibitem[{{Matsuda} et~al.(2020){Matsuda}, {Tanaka}, {Uchida}, {Amano}, and {Tsuru}}]{mtu20}
{Matsuda}, M., {Tanaka}, T., {Uchida}, H., {Amano}, Y., and {Tsuru}, T.~G. (2020).
\newblock {Temporal and spatial variation of synchrotron X-ray stripes in Tycho's supernova remnant}.
\newblock \emph{\pasj} 72, 85.
\newblock \doi{10.1093/pasj/psaa075}
\bibAnnoteFile{mtu20}

\bibitem[{{Mercuri} et~al.(2025){Mercuri}, {Greco}, {Vink}, {Ferrazzoli}, and {Perri}}]{mgv25}
{Mercuri}, A., {Greco}, E., {Vink}, J., {Ferrazzoli}, R., and {Perri}, S. (2025).
\newblock {Revisiting the X-Ray Polarization of the Shell of Cassiopeia A Using Spectropolarimetric Analysis}.
\newblock \emph{\apj} 986, 6.
\newblock \doi{10.3847/1538-4357/adcedb}
\bibAnnoteFile{mgv25}

\bibitem[{{Meshkov}(1969)}]{mes69}
{Meshkov}, E.~E. (1969).
\newblock {Instability of the interface of two gases accelerated by a shock wave}.
\newblock \emph{Fluid Dynamics} 4, 101--104.
\newblock \doi{10.1007/BF01015969}
\bibAnnoteFile{mes69}

\bibitem[{{Miceli} et~al.(2014){Miceli}, {Acero}, {Dubner}, {Decourchelle}, {Orlando}, and {Bocchino}}]{mad14}
{Miceli}, M., {Acero}, F., {Dubner}, G., {Decourchelle}, A., {Orlando}, S., and {Bocchino}, F. (2014).
\newblock {Shock-Cloud Interaction and Particle Acceleration in the Southwestern Limb of SN 1006}.
\newblock \emph{\apjl} 782, L33.
\newblock \doi{10.1088/2041-8205/782/2/L33}
\bibAnnoteFile{mad14}

\bibitem[{{Parizot} et~al.(2006){Parizot}, {Marcowith}, {Ballet}, and {Gallant}}]{pmb06}
{Parizot}, E., {Marcowith}, A., {Ballet}, J., and {Gallant}, Y.~A. (2006).
\newblock {Observational constraints on energetic particle diffusion in young supernovae remnants: amplified magnetic field and maximum energy}.
\newblock \emph{\aap} 453, 387--395.
\newblock \doi{10.1051/0004-6361:20064985}
\bibAnnoteFile{pmb06}

\bibitem[{{Prokhorov} et~al.(2024){Prokhorov}, {Yang}, {Ferrazzoli}, {Vink}, {Slane}, {Costa} et~al.}]{pyf24}
{Prokhorov}, D.~A., {Yang}, Y.-J., {Ferrazzoli}, R., {Vink}, J., {Slane}, P., {Costa}, E., et~al. (2024).
\newblock {Evidence for a shock-compressed magnetic field in the northwestern rim of Vela Jr. from X-ray polarimetry}.
\newblock \emph{\aap} 692, A59.
\newblock \doi{10.1051/0004-6361/202452062}
\bibAnnoteFile{pyf24}

\bibitem[{{Reynolds} et~al.(2007){Reynolds}, {Borkowski}, {Hwang}, {Hughes}, {Badenes}, {Laming} et~al.}]{rbh07}
{Reynolds}, S.~P., {Borkowski}, K.~J., {Hwang}, U., {Hughes}, J.~P., {Badenes}, C., {Laming}, J.~M., et~al. (2007).
\newblock {A Deep Chandra Observation of Kepler's Supernova Remnant: A Type Ia Event with Circumstellar Interaction}.
\newblock \emph{\apjl} 668, L135--L138.
\newblock \doi{10.1086/522830}
\bibAnnoteFile{rbh07}

\bibitem[{{Reynolds} and {Keohane}(1999)}]{rk99}
{Reynolds}, S.~P. and {Keohane}, J.~W. (1999).
\newblock Maximum energies of shock-accelerated electrons in young shell supernova remnants.
\newblock \emph{\apj} 525, 368--374
\bibAnnoteFile{rk99}

\bibitem[{{Reynoso} et~al.(2013){Reynoso}, {Hughes}, and {Moffett}}]{rhm13}
{Reynoso}, E.~M., {Hughes}, J.~P., and {Moffett}, D.~A. (2013).
\newblock {On the Radio Polarization Signature of Efficient and Inefficient Particle Acceleration in Supernova Remnant SN 1006}.
\newblock \emph{\aj} 145, 104.
\newblock \doi{10.1088/0004-6256/145/4/104}
\bibAnnoteFile{rhm13}

\bibitem[{{Richmyer}(1960)}]{ric60}
{Richmyer}, R.~D. (1960).
\newblock {Taylor instability in shock acceleration of compressible fluids}.
\newblock \emph{Communications on Pure and Applied Mathematics} XIII, 297--319.
\newblock \doi{10.1002/cpa.3160130207}
\bibAnnoteFile{ric60}

\bibitem[{{Rosenberg}(1970)}]{ros70}
{Rosenberg}, I. (1970).
\newblock {Distribution of brightness and polarization in Cassiopeia A at 5.0 GHz}.
\newblock \emph{\mnras} 151, 109.
\newblock \doi{10.1093/mnras/151.1.109}
\bibAnnoteFile{ros70}

\bibitem[{{Sapienza} et~al.(2022){Sapienza}, {Miceli}, {Bamba}, {Katsuda}, {Nagayoshi}, {Terada} et~al.}]{smb22}
{Sapienza}, V., {Miceli}, M., {Bamba}, A., {Katsuda}, S., {Nagayoshi}, T., {Terada}, Y., et~al. (2022).
\newblock {A Spatially Resolved Study of Hard X-Ray Emission in Kepler's Supernova Remnant: Indications of Different Regimes of Particle Acceleration}.
\newblock \emph{\apj} 935, 152.
\newblock \doi{10.3847/1538-4357/ac8160}
\bibAnnoteFile{smb22}

\bibitem[{{Sapienza} et~al.(2024){Sapienza}, {Miceli}, {Petruk}, {Bamba}, {Katsuda}, {Orlando} et~al.}]{smp24}
{Sapienza}, V., {Miceli}, M., {Petruk}, O., {Bamba}, A., {Katsuda}, S., {Orlando}, S., et~al. (2024).
\newblock {Time Evolution of the Synchrotron X-Ray Emission in Kepler's Supernova Remnant: The Effects of Turbulence and Shock Velocity}.
\newblock \emph{\apj} 973, 105.
\newblock \doi{10.3847/1538-4357/ad6566}
\bibAnnoteFile{smp24}

\bibitem[{{Toptygin} and {Fleishman}(1987)}]{tf87}
{Toptygin}, I.~N. and {Fleishman}, G.~D. (1987).
\newblock {A Role of Cosmic-Rays in Generation of Radio and Optical Radiation by Plasma Mechanisms}.
\newblock \emph{\apss} 132, 213--248.
\newblock \doi{10.1007/BF00641755}
\bibAnnoteFile{tf87}

\bibitem[{{Tsuji} et~al.(2021){Tsuji}, {Uchiyama}, {Khangulyan}, and {Aharonian}}]{tuk21}
{Tsuji}, N., {Uchiyama}, Y., {Khangulyan}, D., and {Aharonian}, F. (2021).
\newblock {Systematic Study of Acceleration Efficiency in Young Supernova Remnants with Nonthermal X-Ray Observations}.
\newblock \emph{\apj} 907, 117.
\newblock \doi{10.3847/1538-4357/abce65}
\bibAnnoteFile{tuk21}

\bibitem[{{Uchiyama} et~al.(2007){Uchiyama}, {Aharonian}, {Tanaka}, {Takahashi}, and {Maeda}}]{uat07}
{Uchiyama}, Y., {Aharonian}, F.~A., {Tanaka}, T., {Takahashi}, T., and {Maeda}, Y. (2007).
\newblock {Extremely fast acceleration of cosmic rays in a supernova remnant}.
\newblock \emph{\nat} 449, 576--578.
\newblock \doi{10.1038/nature06210}
\bibAnnoteFile{uat07}

\bibitem[{{Vink}(2006)}]{vin06}
{Vink}, J. (2006).
\newblock {X-ray High Resolution and Imaging Spectroscopy of Supernova Remnants}.
\newblock In \emph{The X-ray Universe 2005}, ed. A.~{Wilson}. vol. 604 of \emph{ESA Special Publication}, 319
\bibAnnoteFile{vin06}

\bibitem[{{Vink}(2020)}]{vin20}
{Vink}, J. (2020).
\newblock \emph{{Physics and Evolution of Supernova Remnants}} (Springer).
\newblock \doi{10.1007/978-3-030-55231-2}
\bibAnnoteFile{vin20}

\bibitem[{{Vink} and {Laming}(2003)}]{vl03}
{Vink}, J. and {Laming}, J.~M. (2003).
\newblock {On the Magnetic Fields and Particle Acceleration in Cassiopeia A}.
\newblock \emph{\apj} 584, 758--769.
\newblock \doi{10.1086/345832}
\bibAnnoteFile{vl03}

\bibitem[{{Vink} et~al.(2022){Vink}, {Prokhorov}, {Ferrazzoli}, {Slane}, {Zhou}, {Asakura} et~al.}]{vpf22}
{Vink}, J., {Prokhorov}, D., {Ferrazzoli}, R., {Slane}, P., {Zhou}, P., {Asakura}, K., et~al. (2022).
\newblock {X-ray polarization detection of Cassiopeia A with IXPE}.
\newblock \emph{arXiv e-prints} , arXiv:2206.06713
\bibAnnoteFile{vpf22}

\bibitem[{{Weisskopf} et~al.(2022){Weisskopf}, {Soffitta}, {Baldini}, {Ramsey}, {O'Dell}, {Romani} et~al.}]{IXPE}
{Weisskopf}, M.~C., {Soffitta}, P., {Baldini}, L., {Ramsey}, B.~D., {O'Dell}, S.~L., {Romani}, R.~W., et~al. (2022).
\newblock {The Imaging X-Ray Polarimetry Explorer (IXPE): Pre-Launch}.
\newblock \emph{Journal of Astronomical Telescopes, Instruments, and Systems} 8, 026002.
\newblock \doi{10.1117/1.JATIS.8.2.026002}
\bibAnnoteFile{IXPE}

\bibitem[{{Williams} et~al.(2016){Williams}, {Chomiuk}, {Hewitt}, {Blondin}, {Borkowski}, {Ghavamian} et~al.}]{wch16}
{Williams}, B.~J., {Chomiuk}, L., {Hewitt}, J.~W., {Blondin}, J.~M., {Borkowski}, K.~J., {Ghavamian}, P., et~al. (2016).
\newblock {An X-Ray and Radio Study of the Varying Expansion Velocities in Tycho{\textquoteright}s Supernova Remnant}.
\newblock \emph{\apjl} 823, L32.
\newblock \doi{10.3847/2041-8205/823/2/L32}
\bibAnnoteFile{wch16}

\bibitem[{{Xu} et~al.(2007){Xu}, {Han}, {Sun}, {Reich}, {Xiao}, {Reich} et~al.}]{xhs07}
{Xu}, J.~W., {Han}, J.~L., {Sun}, X.~H., {Reich}, W., {Xiao}, L., {Reich}, P., et~al. (2007).
\newblock {Polarization observations of SNR G156.2+5.7 at {\ensuremath{\lambda}}6 cm}.
\newblock \emph{\aap} 470, 969--975.
\newblock \doi{10.1051/0004-6361:20077549}
\bibAnnoteFile{xhs07}

\bibitem[{{Zhou} et~al.(2023){Zhou}, {Prokhorov}, {Ferrazzoli}, {Yang}, {Slane}, {Vink} et~al.}]{zpf23}
{Zhou}, P., {Prokhorov}, D., {Ferrazzoli}, R., {Yang}, Y.-J., {Slane}, P., {Vink}, J., et~al. (2023).
\newblock {Magnetic Structures and Turbulence in SN 1006 Revealed with Imaging X-Ray Polarimetry}.
\newblock \emph{\apj} 957, 55.
\newblock \doi{10.3847/1538-4357/acf3e6}
\bibAnnoteFile{zpf23}

\bibitem[{{Zhou} et~al.(2025){Zhou}, {Slane}, {Prokhorov}, {Vink}, {Ferrazzoli}, {Cotton} et~al.}]{zsp25}
{Zhou}, P., {Slane}, P., {Prokhorov}, D., {Vink}, J., {Ferrazzoli}, R., {Cotton}, W., et~al. (2025).
\newblock {X-Ray Polarization in SN 1006 Southwest Shows Spatial Variations and Differences in the Radio Band}.
\newblock \emph{\apj} 986, 210.
\newblock \doi{10.3847/1538-4357/add532}
\bibAnnoteFile{zsp25}

\bibitem[{{Zirakashvili} and {Aharonian}(2007)}]{za07}
{Zirakashvili}, V.~N. and {Aharonian}, F. (2007).
\newblock {Analytical solutions for energy spectra of electrons accelerated by nonrelativistic shock-waves in shell type supernova remnants}.
\newblock \emph{\aap} 465, 695--702.
\newblock \doi{10.1051/0004-6361:20066494}
\bibAnnoteFile{za07}

\end{thebibliography}

\end{document}